\documentclass[5p]{elsarticle}

\usepackage{amsmath}
\usepackage{color}
\usepackage{graphicx}
\usepackage[colorlinks,linkcolor=red,anchorcolor=blue,citecolor=green]{hyperref}
\usepackage[ruled]{algorithm2e}
\usepackage{bm}
\usepackage{natbib}
\usepackage{booktabs}
\usepackage{tabularx}
\usepackage{multirow}
\usepackage{multicol}
\usepackage{setspace}
\journal{Optics $\&$ Lasers in Engineering}
\newcommand{\linu}{\underline{\hspace{0.5em}}}
\usepackage{flushend}
\title{Microscopic 3D measurement of shiny surfaces based on a multi-frequency phase-shifting scheme}
\begin{document}
\begin{frontmatter}
        \title{Microscopic 3D measurement of shiny surfaces based on a multi-frequency phase-shifting scheme}
        \author{Yan Hu,$^{a,b}$ Qian Chen,$^{a,\dag}$ Yichao Liang,$^{a,b}$ Shijie Feng,$^{a,b}$ Tianyang Tao,$^{a,b}$ Chao Zuo$^{a,b,*}$}
        \address{
        $^a$Jiangsu Key Laboratory of Spectral Imaging $\&$ Intelligent Sense, Nanjing University of Science and Technology, Nanjing, Jiangsu Province 210094, China\\
        $^b$Smart Computational Imaging (SCI) Laboratory, Nanjing University of Science and Technology, Nanjing, Jiangsu Province 210094, China\\
        $^{\dag}$chenqian@njust.edu.cn\\
        $^{*}$zuochao@njust.edu.cn
        }

        \begin{abstract}
       Microscopic fringe projection profilometry is a powerful 3D measurement technique with a theoretical measurement accuracy better than one micron provided that the measured targets can be imaged with good fringe visibility. However, practically, the 3D shape of the measured surface can hardly be fully reconstructed due to the defocus of the dense fringes and complex surface reflexivity characteristics, which lead to low fringe quality and intensity saturation. To address this problem, we propose to calculate phases of these highlighted areas from a subset of the fringe sequence which is not subjected to the intensity saturation. By using the proposed multi-frequency phase-shifting scheme, the integrity of the 3D surface reconstruction can be significantly improved. The ultimate phase maps obtained from unsaturated intensities are used to achieve high-accuracy 3D recovering of shiny surfaces based on a phase stereo matching method. Experimental results on different metal surfaces show that our approach is able to retrieve the complete morphology of shiny surfaces with high accuracy and fidelity.
        \end{abstract}

        \end{frontmatter}

 \section{Introduction} \noindent
        The principle of structured light and triangulation has been widely used in the range of 3D optical metrology applications \cite{lazaros_review_2008}. Generally, periodic sinusoidal fringe patterns are projected onto an inspected object, and the fringe pattern is distorted by the modulation of the object. To quantitatively calculate the amount of the modulation and reconstruct the 3D result of the target, the phase value coded in the fringe pattern needs to be accurately retrieved. For now, two commonly used phase retrieval algorithms are Fourier transform based algorithms \cite{su2010dynamic, zuo2018micro, hu2018dynamic} and phase-shifting based algorithms \cite{zuo2013high, zuo2018phase, feng2018robust, hu2017absolute}. Fourier transform based algorithms are commonly used in dynamic measurement while phase-shifting based algorithms are more suitable for high-accuracy measurement owning to its pixel independently mathematical operational nature. Our recent work has shown that by the phase-based stereo matching method, the essential nonlinear response function of the digital projector can be neglected because the phase errors in different views are automatically balanced out \cite{hu2018microscopic}. However, the phase-based stereo matching method prone to fail when dealing with the objects with shiny surfaces. The integrity of the reconstructed model is affected by the highlight regions because the phase in these areas cannot be calculated by dense fringe images.

        Shiny surfaces are highly reflective, and thus the light intensity cannot be transformed linearly because of the limited dynamic range of digital cameras. One of the state-of-the-art techniques for this situation is called the high-dynamic range 3D shape measurement \cite{feng2018high}, which can be classified into two categories: equipment-based techniques and algorithm-based techniques. For the group of the equipment-based techniques, optimal parameters of the equipment, e.g., the exposure time of the camera \cite{zhang2009high, ekstrand2011autoexposure, zhong2015enhanced} or the projector \cite{waddington2010saturation, babaie2015dynamics, li2014adaptive, lin2016adaptive} are desired to help capture visible fringe at both shiny and dark surfaces. Additional optical based methods, e.g., using a polarizer to scan shiny objects have also been investigated \cite{chen2007polarization, salahieh2014multi}, based on which the polarized highlight intensity can be effectively suppressed. Also, there are hybrid methods by modifying camera exposure, but also taking into account strategies of introducing additional equipment, changing the viewing position, or adjusting parameters of projectors to capture HDR images \cite{feng2014general, liu20113d, jiang2012high}. Based on the maximum intensity modulation, a fast HDR solution employing a high-speed projector to project intensity-varying fringe images at 700 Hz is proposed \cite{zhao2014rapid}.

        For shiny surfaces, however, the problem of saturation may not be readily handled by merely decreasing the exposure time or the intensity of the projected light in some cases. Thus, algorithm-based techniques are also developed, which mainly rely on well-designed algorithms to extract phase values from raw fringe images when free adjustment of the camera or the projector is not allowed, or additional equipment is not available. Yin et al. \cite{yin2017high} suggested measuring shiny surfaces with a single color image. Alternatively, Jiang et al. \cite{jiang2016high} proposed a real-time HDR 3-D scanning method by projecting additional inverted fringe patterns. Chen et al. \cite{chen2008phase} found that the phase-shifting methods can overcome the image saturation if the number of the phase shift is high enough so that at least three unsaturated fringe intensities can be recorded successfully. Moreover, Chen et al. \cite{chen2016high} proposed a technique, by which the phase is also calculated from raw phase-shifting images but without considering whether they are saturated or not. Hu et al. \cite{hu2010study, hu2010further} introduced a phase-shifting based method by taking advantage of no less than three unsaturated fringe image from standard $N$-step phase-shifting algorithm.

        These related works have successfully addressed HDR measurement problems by various means. But when it comes to microscopic imaging, those shiny parts illuminated by black stripes can no longer be imaged purely black but will be affected by the white stripes due to the short depth of field of the microscopic projection system. In this case, there will be more saturated areas when using higher frequency fringes. Inspired by the works proposed by Hu et al. \cite{hu2010study, hu2010further}, we propose an HDR measurement method which takes advantage of the generalized phase-shifting algorithm and considers the spatial frequency characteristic of multi-frequency fringe patterns in the microscopic measurement applications. Specifically, phase values in the regions without any saturation are calculated directly using the standard phase-shifting algorithm. In those partially saturated regions, the generalized phase-shifting algorithm is used to calculate the wrapped phase. For those over-saturation areas with less than three unsaturated intensities, the phases probably retrievable from lower frequency fringe images are used to fill up the phase map in order to increase the measurement integrity. After phase unwrapping and stereo matching of the dual-view telecentric measurement system \cite{hu2018microscopic}, high-accuracy 3D reconstruction of shiny surfaces (HDR objects) can be successfully achieved. The experiments demonstrate that the proposed multi-frequency phase-shifting scheme accommodates the measurements of various kinds of shiny objects with the measurement high-accuracy within one micron.

    \section{Principals}
        \subsection{{Generalized phase-shifting algorithm}} 
\begin{figure}[htbp]
            \centerline{\includegraphics[width=1.0\columnwidth]{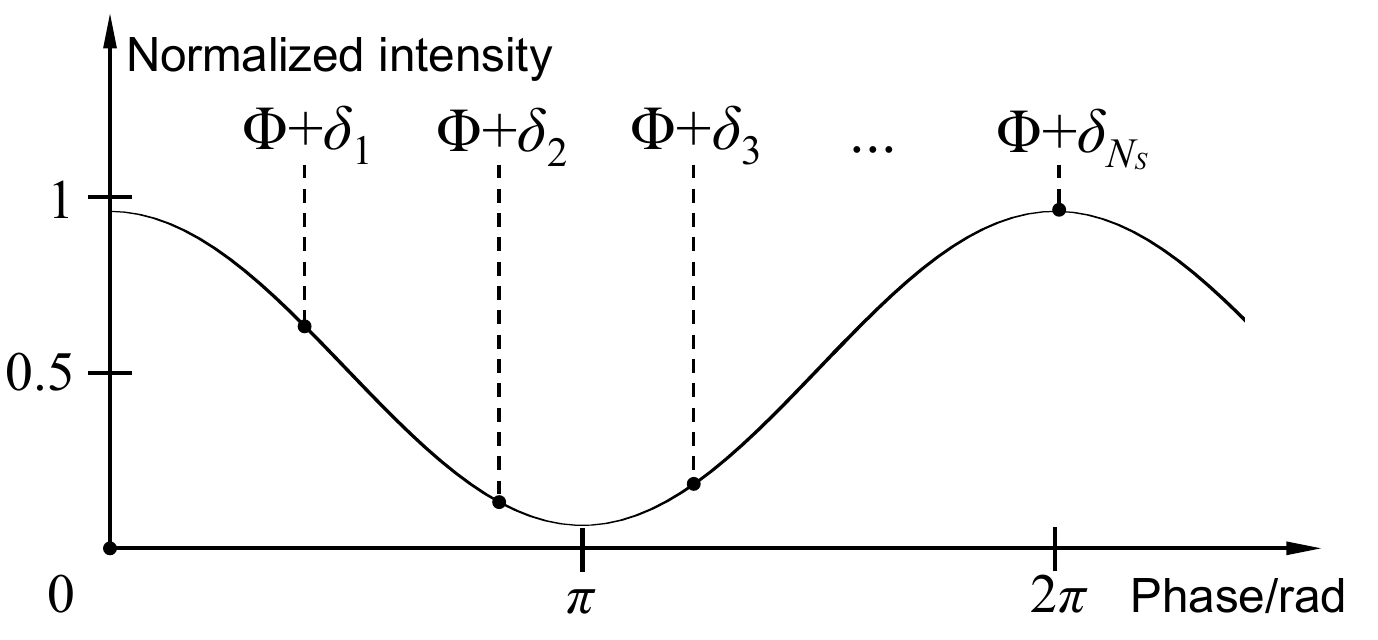}}
            \caption{
                    Sketch map of the relation between the intensity and the shifted phase in a phase-shifting process.
                    }
            \label{fig1}
         \end{figure}

\noindent
        Traditional approaches to generate fringe images of fringe projection profilometry involve laser interferometry, physical grating, or slide projector. However, with recent developments in the area of the digital display, digital projectors have been increasingly applied as the projection units. Based on the controllable phase-shifting amount, the recorded fringe image with $\delta_n$ phase-shifting can be expressed by
        \begin{equation}\label{2.1}
             I_n(u,v)=I_0(u,v)\{1+\alpha(u,v)\cos[\Phi(u,v) + \delta_n],
        \end{equation}
        where $(u,v)$ is the pixel coordinate of the camera, ${I}_{0}$ is the average intensity, $\alpha$ is the fringe contrast, $\Phi$ is the phase distribution to be measured. $\delta_n$ is the shifted reference phase ($n=1,...,N$). The phase distribution $\Phi$ can be calculated independently over no less than three phase-shifted intensities as shown in Fig. \ref{fig1}. Based on minimizing a criterion concerning the difference between ideal intensities and captured intensities \cite{lai1991generalized}, we can obtain the wrapped phase $\phi$ corresponding to $\Phi$ as
        \begin{equation}\label{2.2}
            \phi = -\arctan \left( \frac{{{\alpha }_{2}}}{{{\alpha }_{1}}} \right),
        \end{equation}
        with
        \begin{equation}\label{2.3}
        \left\{ \begin{array}{l}
        {{\alpha }_{1}}={{c}_{21}}\sum\limits_{n=1}^{N}{{{I}_{n}}}+{{c}_{22}}\sum\limits_{n=1}^{N}{{{I}_{n}}\cos \left( {{\delta }_{n}} \right)}+{{c}_{23}}\sum\limits_{n=1}^{N}{{{I}_{n}}\sin \left( {{\delta }_{n}} \right)} \\
        {{\alpha }_{2}}={{c}_{31}}\sum\limits_{n=1}^{N}{{{I}_{n}}}+{{c}_{32}}\sum\limits_{n=1}^{N}{{{I}_{n}}\cos \left( {{\delta }_{n}} \right)}+{{c}_{33}}\sum\limits_{n=1}^{N}{{{I}_{n}}\sin \left( {{\delta }_{n}} \right)} \\
        \end{array} \right.,
        \end{equation}
        and the coefficients $c_{ij,~i=2,3;~j=1,2,3}$ in equation (\ref{2.3}) can be calculated though
        \begin{equation}\label{2.4}
        \begin{array}{l}
          \textbf{C}=\left[ \begin{array}{ccc}
           {{c}_{11}} & {{c}_{12}} & {{c}_{13}}  \\
           {{c}_{21}} & {{c}_{22}} & {{c}_{23}}  \\
           {{c}_{31}} & {{c}_{32}} & {{c}_{33}}  \\
        \end{array} \right]={\textbf{A}^{-1}} \\
         ={{\left[ \begin{array}{lll}
           N & \sum\limits_{n=1}^{N}{\cos \left( {{\delta}_{n}} \right)} & \sum\limits_{n=1}^{N}{\sin \left( {{\delta }_{n}} \right)}  \\
           \sum\limits_{n=1}^{N}{\cos \left( {{\delta}_{n}} \right)} & \sum\limits_{n=1}^{N}{{{\cos }^{2}}\left( {{\delta }_{n}} \right)} & \sum\limits_{n=1}^{N}{\sin \left( {{\delta }_{n}} \right)\cos \left( {{\delta}_{n}} \right)}  \\
           \sum\limits_{n=1}^{N}{\sin \left( {{\delta}_{n}} \right)} & \sum\limits_{n=1}^{N}{\sin \left( {{\delta }_{n}} \right)\cos \left( {{\delta}_{n}} \right)} & \sum\limits_{n=1}^{N}{{{\sin}^{2}}\left( {{\delta}_{n}} \right)}  \\
        \end{array} \right]}^{-1}.}
        \end{array}
        \end{equation}
        Since the phase step $\delta_n$ is strictly controlled, two-dimensional wrapped phase distribution $\phi(u,v)$ can be obtained from equations (\ref{2.2}) - (\ref{2.4}). Particularly, if $\delta_n$ is equally divided by an integer $N_S$ in the range $[0,2\pi)$, equations (\ref{2.2}) - (\ref{2.4}) can be simplified as the standard phase-shifting algorithm:
        \begin{equation}\label{2.5}
            \phi = -\arctan \left[ \frac{ \sum\nolimits_{n=1}^{N_S}{{I}_{n}}\sin({{\delta }_{n}})}{\sum\nolimits_{n=1}^{N_S}{{I}_{n}}\cos({{\delta }_{n}})}\right].
        \end{equation}

        \subsection{{Multi-frequency phase-shifting scheme for HDR surface measurement}} \noindent
        The accuracy of the phase values depends on the phase-shifting step number and the fringe contrast. When the final absolute phase is scaled into the same range $[0,2\pi)$, the phase error variance can be stated as \cite{Zuo2016Temp}.
        \begin{equation}\label{2.6}
            \sigma^2_\Phi = \frac{2\sigma^2}{N_Sf^2B^2}.
        \end{equation}
        Here, $\sigma$ is the variance of a Gaussian distributed additive noise. $N_S$ is the phase-shifting step number. $f$ is the fringe frequency, indicating the fringe density. $B$ is the fringe modulation. If the phase-shifts are confirmed, in order to acquire a higher phase accuracy, we should use patterns with higher frequency $(f)$ and try to capture images with better fringe visibility $(B)$ as well.

       \begin{figure}[htbp]
            \centerline{\includegraphics[width=0.9\columnwidth]{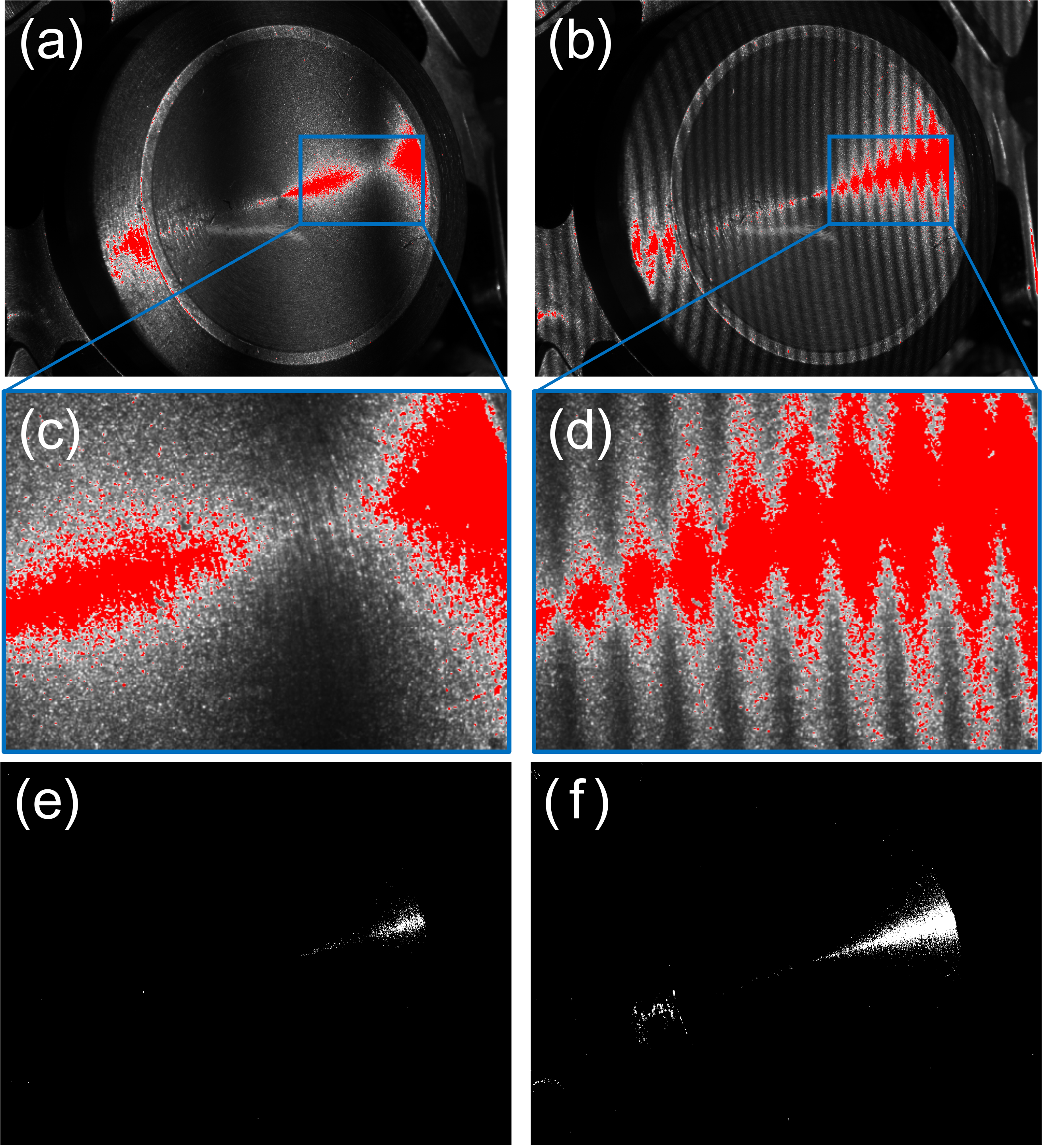}}
            \caption{
                    Comparison of the saturation degree between fringes with different frequency. (a) and (b) Raw fringe images; (c) and (d) Magnified details of saturated parts; (e) and (f) Indexes that show regions with less than three unsaturated intensities.}
            \label{fig2}
         \end{figure}

        However, in the projection system, the low depth of field leads to a significantly attenuated fringe contrast when increasing the fringe density. Furthermore, low contrast fringes falling on the shining surface can easily cause intensity saturation. An example in Fig. \ref{fig2} shows a metal surface covered by fringe patterns with two different frequencies. The periods the fringe pattern in Fig. \ref{fig2}(a) and (b) are 144 and 12, respectively. Due to the slight defocus caused by the shallow depth of field, the dark stripe in the marked sub-region in Fig. \ref{fig2}(b) is polluted by the nearby non-black light and thus saturation occurs, as presented in the magnified details in Fig. \ref{fig2}(d). However, when this region is projected by a less dense fringe as shown in Fig. \ref{fig2}(a), the dark stripe gives more unsaturated pixels, as Fig. \ref{fig2}(c) presents, and thus the phase values in such regions can possibly be retrieved.

        To intuitively show the difference, we extract the region with less than three unsaturated intensities. The white areas in Figs. \ref{fig2}(e) and (f) indicate the pixels with less than three unsaturated intensities. In this kind of area, the phase cannot be calculated because the unknowns are more than the conditions. Clearly, Fig. \ref{fig2}(e) has quite fewer pixels where the phase cannot be calculated.

        When adjusting the system parameters, we should maintain the saturated region in a small part. We do not decrease the exposure time of the whole field only to decrease the saturation area. However, for those samples with a large proportion of saturation, the right way is decreasing the exposure time to leave a small region with saturation. In traditional multi-frequency phase-shifting methods, the fringe patterns with lower frequency are used to provide a reference phase map for phase unwrapping, so that the ultimate measurement result is determined by the fringe images with the highest frequency. Actually, in the saturated regions, the ultimate phase values can be replaced by that derived from the less dense fringe images with less saturated intensities. In this way, the 3D reconstruction can be preserved as complete as possible. For this purpose, we propose a multi-frequency fringe based scheme for HDR surface measurement. Three steps, including image data preparation, saturation detection and compensation algorithms, and phase stereo matching allow the high-accuracy microscopic 3D measurement of shiny surfaces.

        $\textbf{Step one}$ is the image data preparation stage, which contains image acquisition, image rectification, and classification according to the fringe frequency. The fringe patterns are sequentially projected with trigger signals for the camera synchronization. $\textbf{Step two}$  is the main part corresponding to the proposed multi-frequency fringe based scheme, which is to calculate the unwrapped phase map by three algorithms. Their definitions are detailed in algorithm \ref {Alg1}, \ref {Alg2}, and \ref {Alg3}, respectively. $\textbf{Step three}$ is phase stereo matching and 3D reconstruction, which will be discussed in the next subsection. Tab. \ref {tab1} lists the description of the variables used in Step two:

        \begin{table}[htbp]
            \centering
            \caption{Information on used patterns to calculate phase maps}
            \begin{tabular}{cp{6.5cm}}
                \toprule[1pt]
                \multicolumn{1}{c}{Items} & \multicolumn{1}{c}{description} \\
                \midrule
                $I^m_{set}$ & the $m^{th}$ fringe image set containing a group of standard phase-shifted fringe images.\\
                $M$ & the total number of the fringe image sets.\\
                $sat{\linu}map^m$ & the pixel-wise map containing the saturated intensity number of $I^m_{set}$.\\
                $N^m_S$ & the phase-shifting number of the $m^{th}$ image set.\\
                $per^m$ & the fringe period corresponding to the $m^{th}$ image set.\\
                $\Phi_{eq}^m$ & equivalent unwrapped phase map of $\Phi^m$.\\
                $k^m$ & the fringe order for phase unwrapping.\\
                $sat{\linu}thr$ & the saturation threshold intensity.\\
                $I\linu slot$ & a temporary storage for the unsaturated $I_n$ in a phase-shifting process.\\
                $\delta\linu slot$ & a temporary storage for $\delta_n$ of unsaturated intensities $I_n$.\\
                $\delta_{std}$ & an array containing the standard phase-shifts.\\
                $ind^m$ & a two-dimensional index map, in which '1' indicates over-saturation.\\
                $\sim ind^m$ & the unary complement of $ind^m$.\\
                ${ind}\linu{rep}^m$ & an index map storing which pixels in $\Phi$ are to be replaced by $\Phi_{eq}^m$.\\
                \bottomrule[1pt]
            \end{tabular}
            \label{tab1}
        \end{table}

        \begin{figure}[htbp]
            \centerline{\includegraphics[width=1.0\columnwidth]{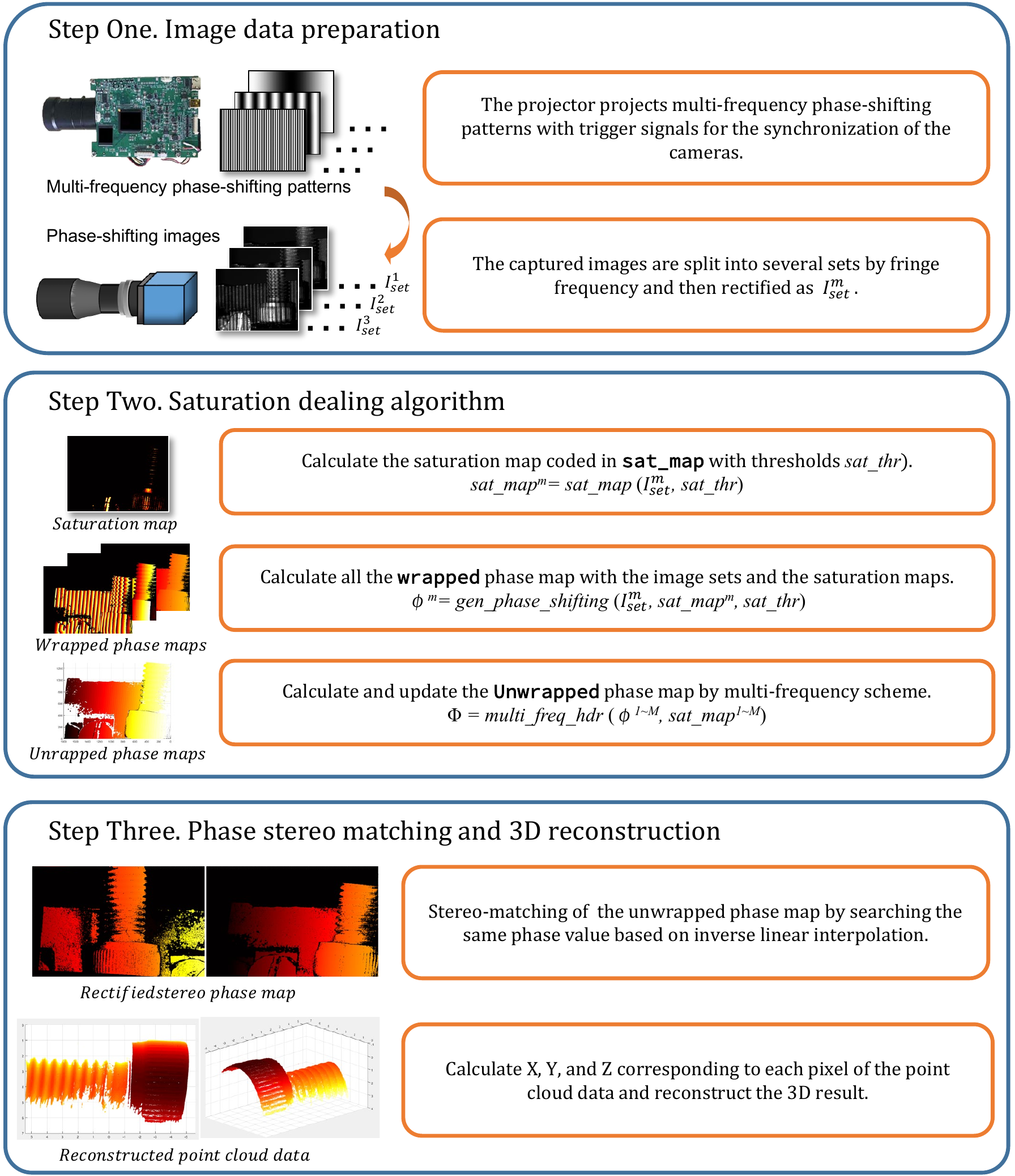}}
            \caption{
                    The flowchart of the proposed algorithm for shiny surface measurement.}
            \label{fig3}
        \end{figure}

        \begin{algorithm}[htbp]
            \caption{$\bf{sat{\linu}map}$}
            \label{Alg1}
            \LinesNumbered
            \KwIn{$I^m_{set}$.
            }
            \KwOut{$sat{\linu}map^m$.\\
            }
            \For{pixel (u,v)}
            {
                \For{i = 1 : $N^m_S$}
                {
                    $jud$ = the i$^{th}$ intensity at pixel $(u,v)$\;
                　　\If {jud $\geq$ sat{\linu}thr}
                    {
                        $sat{\linu}map^m(u,v)=sat{\linu}map^m(u,v)+1$\;
                    }
                }
            }
        \end{algorithm}

        \begin{algorithm}[htbp]
            \caption{$\bf{gen{\linu}phase{\linu}shifting}$}
            \label{Alg2}
            \LinesNumbered
            \KwIn{$I^m_{set}$, $sat{\linu}map^m$. \\
            }
            \KwOut{$wrapped~phase ~\phi^m$.\\
            }
            \For{pixel (u,v)}
            {
                \eIf{$sat{\linu}map^m(u,v) = 0$}
                {
                    $\phi^m(u,v)$ = Eq. (\ref{2.5}).
                }
                {
                    $I{\linu}slot = I^{1\sim N_S}_{set}(u ,v)$\;
                    $\delta\linu slot = \delta_{1\sim N_S}$\;
                    $index = {(I{\linu}slot \geq sat{\linu}thr)}$;\\
                    $I{\linu}slot (index) = empty$\;
                    $\delta\linu slot (index) = empty$\;
                    \eIf{$length(I{\linu}slot)<3$}
                    {
                        \textbf{continue}\;
                    }
                    {
                        $\phi^m(u,v)$ = Eq. (\ref{2.2});
                    }
                }
            }
        \end{algorithm}

        \begin{algorithm}[htbp]
            \caption{$\bf{multi{\linu}freq{\linu}hdr}$}
            \label{Alg3}
            \LinesNumbered
            \KwIn{$sat{\linu}map^{1\sim M}$, $\phi^{1\sim M}$, $k^m$. \\
            }
            \KwOut{$Unwrapped~phase~\Phi$.\\
            }
            $\Phi^1 = \phi^1$\;
            \For{ m = 2 : $M$}
            {
                $k^m = \text{round}[(\Phi^{m-1}\cdot per^{m-1}/per^m-\phi^m)/2\pi]$\;
                $\Phi^m = \phi^m + 2\pi k^m$;
            }
            \For{ m = 1 : $M$}
            {
                $\Phi^m_{eq} = \Phi^m \cdot per^m/per^M$;
            }
            \For{ m = 1 : $M$}
            {
                $ind^m = sat{\linu}map^m > (N_S - 3)$;
            }
            \For{ m = 1 : $M$}
            {
                ${ind}\linu{rep}^m = ind^M~\&~ind^{(M-1)}~\cdots ~\& \sim{ind^m}$;
            }
            \For{ m = 1 : $M$}
            {
                $\Phi({ind}\linu{rep}^m) = \Phi^m_{eq}({ind}\linu{rep}^m)$;
            }
            \end{algorithm}

        Algorithm \ref {Alg1} is to count the number of saturated intensities at each pixel in an image set by $\textbf{sat{\linu}map}$. The stored information is to be referred in the phase unwrapping stage with the saturation levels of different fringe periods being considered. Algorithm \ref {Alg2} is the phase calculation algorithm for the partially saturated phase-shifting fringe images by $\textbf{gen{\linu}phase{\linu}shifting}$. Invalid intensities at each pixel are eliminated and the generalized phase-shifting algorithm corresponding to equations (\ref{2.2}) - (\ref{2.4}) is applied for the phase calculation. Algorithm \ref {Alg3} is the automatic fusion method for the correctness of the unwrapped phase by $\textbf{multi{\linu}freq{\linu}hdr}$. As shown in Figs. \ref{fig2}(e) and (f), because of the defocus of the projected pattern, the denser fringes are easier to be blurred and thus there will be fewer pixels with no less than three valid intensities. For these pixels, we use the equivalent phase derived from the less dense fringe set to fill up the unwrapped phase map. This algorithm searches the phase candidates from the relatively denser image set and then the looser ones for better noise immunity.

        To make it easier to understand the whole process of the proposed scheme, we draw a flowchart containing the three steps as shown in Fig. \ref{fig3}, in which a screw thread is measured as an example.

    \subsection{{3D reconstruction based on phase stereo matching}} 
\begin{figure}[htbp]
            \centerline{\includegraphics[width=1.0\columnwidth]{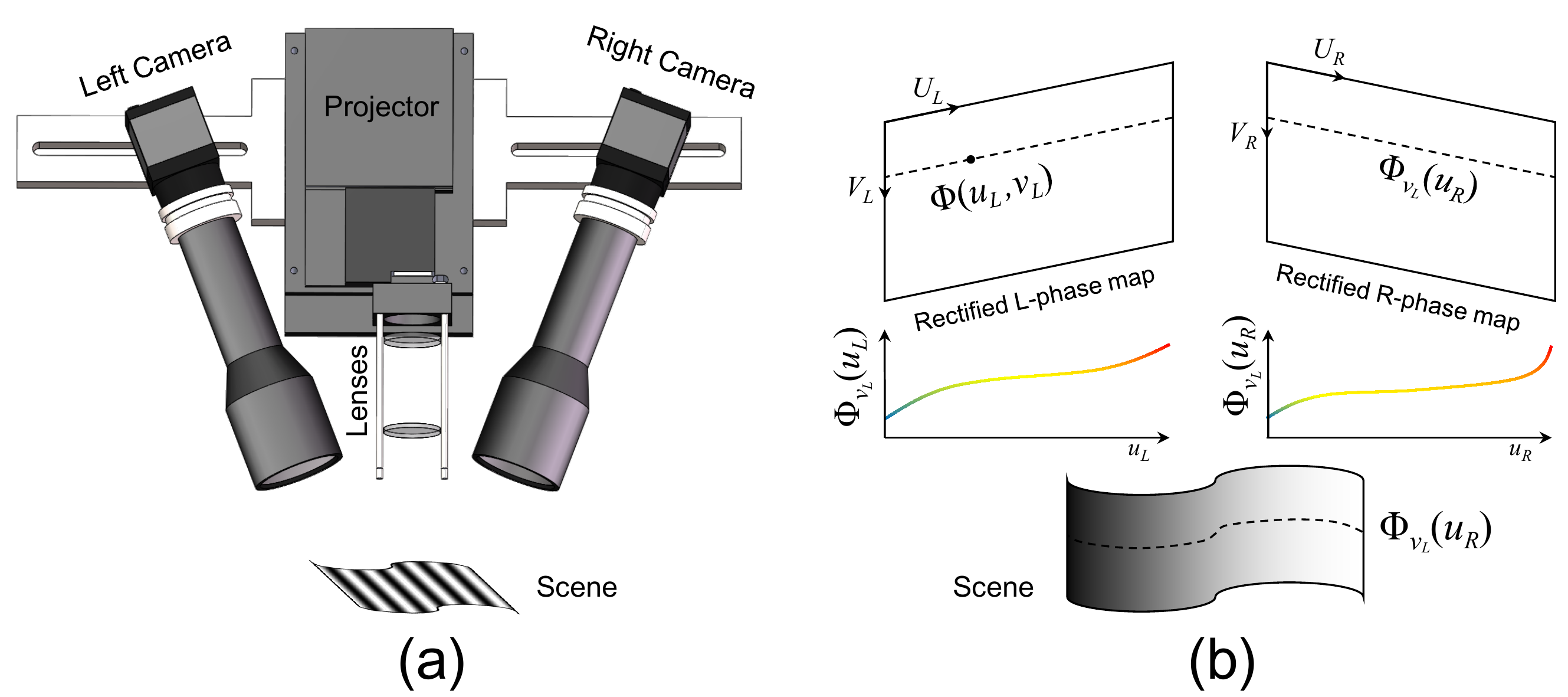}}
            \caption{
            (a) Simplified structure model of the system; (b) Illustration of the bilocular matching based on the unwrapped phase map.
            }
            \label{fig4}
         \end{figure}

\noindent
        The structure model of our measurement system is presented in Fig. \ref{fig4}(a). Sinusoidal patterns encoded with horizontally increased phase maps are projected in sequence from the digital projector. The fringes are deformed by the object and then captured by two telecentric cameras. The camera model is acA2040-120um with the pixel size of 3.45 $\mu$m, and resolution of 2048$\times$1536. The model of the telecentric lens is XF-UTL-0296X175 with a magnification of 0.296$\times$, depth of field of 16.1 mm, and spatial resolution of 31.2$\mu$m. In the experiments, $M$ is 4, $N^{1\sim4}_S$ are [912 144 24 12], and $Per^{\sim}_S$ are [912 144 24 12]. The projection speed is 40 frames per second, which means a measurement only takes 1.2 seconds. By using the proposed multi-frequency fringe based method, the absolute phase value $\Phi$ from both cameras can be obtained for the stereo matching.

        Telecentric epipolar rectification is performed on the fringe images first. Without loss of generality, the left camera is considered as the main camera. As Fig. \ref{fig4}(b) shows, for a pixel $(u_L,v_L)$ on the left camera with phase value $\Phi(u_L,v_L)$, the task is to find the corresponding pixel $u_R$ in the ${v_L}^{th}$ row on the right image. Because the fringe direction is vertical so that the unwrapped phase value increases along the horizontal direction. The integral pixel $u^I_R$ that has the nearest phase value to $\Phi(u_L,v_L)$ in the ${v_L}^{th}$ row is firstly obtained with its phase being $\Phi({u^I_R})$. Then sub-pixel coordinate ${u_R}$ is thereby calculated based on inverse linear interpolation:
            \begin{equation}\label{2.7}
            {u_{R}}=u^{I}_{R}+ \left\{\begin{array}{l} \dfrac{{{\Phi}}({u_L},{v_L})-{{\Phi}}({u^I_R})}{{{\Phi}}({u^I_R}+1)-{{\Phi}}({u^I_R})},{{\Phi}}(u_L,v_L)>\Phi(u^I_{R})\\[15pt]
            \dfrac{{{\Phi }}({u_L},{v_L})-{{\Phi}}({u^I_R})}{{{\Phi }}(u^I_{R})-{{\Phi }}(u^I_{R}-1)},{{\Phi }}(u_L,v_L)\le\Phi(u^I_{R}) \\
            \end{array}\right..
            \end{equation}
        After completing the stereo matching, we have the matched pixel pairs. For more accurate sub-pixel searching, ${u_R}$ can be interpolated with a more complex fitting method that involves more neighboring pixels but more time will be consumed. Together with the new camera parameters after the epipolar rectification, the point cloud data can be directly derived \cite{hu2018microscopic}.

         \begin{figure}[htbp]
            \centerline{\includegraphics[width=1.0\columnwidth]{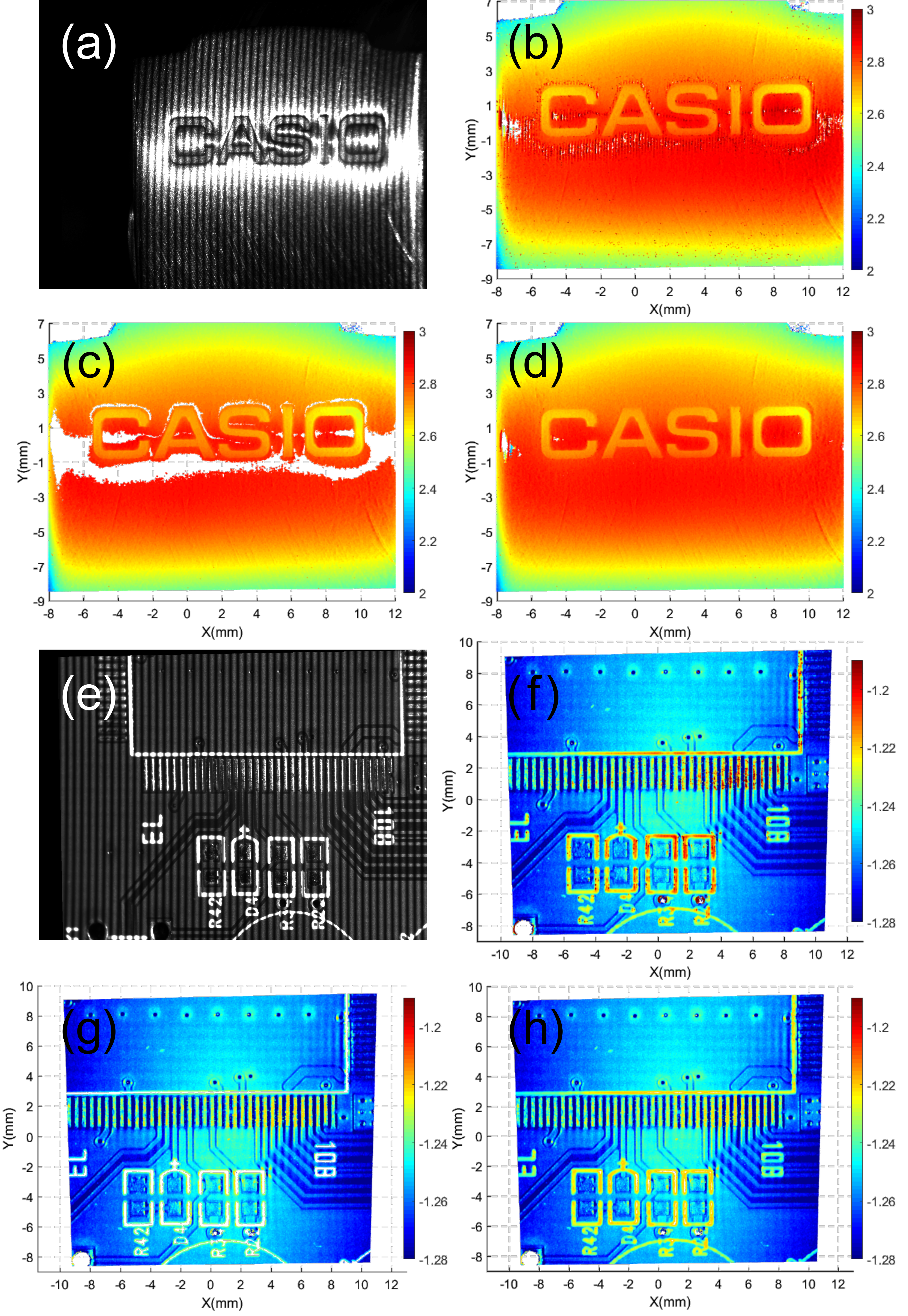}}
            \caption{Comparison of our method and the traditional method when dealing with partially saturated targets. (a) Fringe image of a stamped logo on a metal watch strap; (e) Fringe image of a printed circuit board; (b) and (f) Results with error and ripples; (c) and (g) Results with error-contained regions eliminated; (d) and (h) Results from our proposed multi-frequency fringe based method for HDR surface.
            }
            \label{fig5}
         \end{figure}
    \section{Experiments}
        \subsection{{comparison with traditional multi-frequency method}} \noindent
        Phase-shifting algorithms assume that the captured intensities vary in the form of a sinusoidal wave as the phase shifts linearly. If the sample is very reflective, a portion of saturated intensities (255 if the image sampling resolution is 8 bit) will replace those intensities larger than the maximum of the sampling limit, which is defined as partial saturation. To compare our proposed method with the traditional multi-frequency method when dealing with partially saturated targets, we conducted measurements of two samples with a shiny surface.

        Fig. \ref{fig5} presents the measurement results. The first sample is a stamped logo on a metal watch strap. As shown in Fig. \ref{fig5}(a), the shiny surface is imaged with quite severe saturation around the letters when projected by fringes. The other sample in Fig. \ref{fig5}(e) is a printed circuit board with metal bonding pad and white silkscreen letters and lines. If the traditional phase-shifting algorithm is applied by using these saturated intensities for the phase calculation, saturated regions on the samples will be incorrectly reconstructed, as shown in Figs. \ref{fig5}(b) and (f). Ripples appear in those partially and totally saturated regions. For those regions with unsaturated intensity number less than three, the phase value can no longer be retrieved, which causes incomplete results, as shown in Figs. \ref{fig5}(c) and (g). By using our proposed multi-frequency fringe based method, we can anyhow acquire a phase map as complete as possible with correct phase values and the 3D surface profile of the shiny samples can thus be reconstructed as well, as shown in Figs. \ref{fig5}(d) and (h).

        Although denser fringe gives lower fringe contrast, they are still preferred because of their high immunity to noise, as equation (\ref{2.6}) manifests. It should also be noted that the success of the proposed method is under a necessary condition, that is, the method should be used in a microscopic 3D measurement system because it is the low depth of field of the microscopic projection system that makes the denser fringe easily defocused. For the measurement system for relatively large scale objects, the field of view of the optical system will be quite bigger and the depth of field is much deeper, thus the fringe contrast nearly keeps unchanged, in which case our method may no longer be valid.

       \subsection{{Measurement of shiny samples}} 
\begin{figure}[htbp]
            \centerline{\includegraphics[width=1.0\columnwidth]{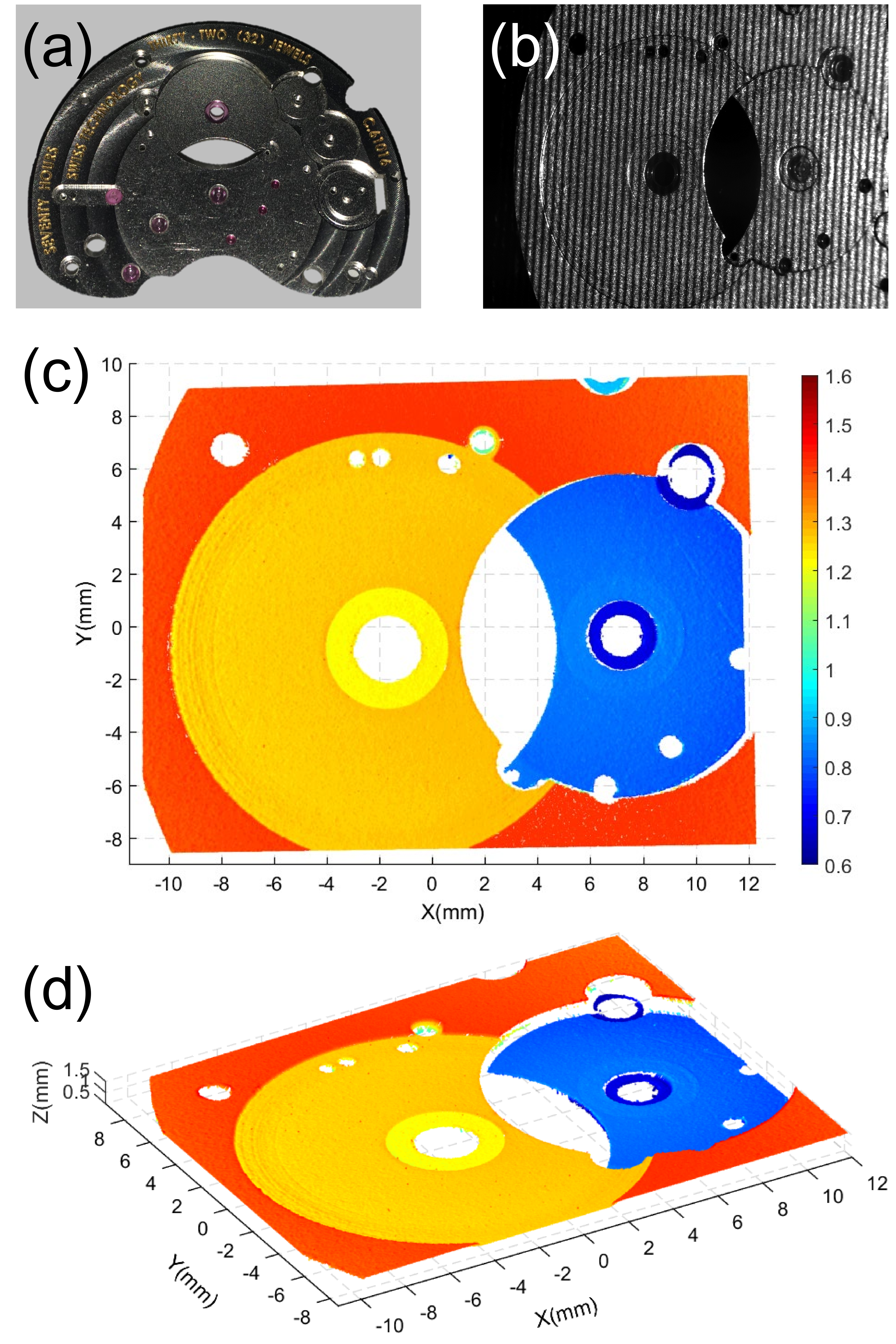}}
            \caption{
                    Experiments on a nickel-plated plate of a mechanical watch. (a) Sample image; (b) One of the fringe image; (c) and (d) Top view and oblique view of the reconstructed 3D model.
            }
            \label{fig6}
         \end{figure}

\noindent
        In order to demonstrate the performance of the proposed method for microscopic measurement of shiny surfaces, we conducted two experiments on metal samples. The first example is a plate of a mechanical watch, which is nickel-plated as shown in Fig. \ref{fig6}(a). We measured its backside and one of the fringe image from the right camera is presented in Fig. \ref{fig6}(b). Due to the high reflection of the metal surface and the weak depth of field of the projected fringe, saturated regions are quite conspicuous. By the processes as drawn in the flowchart in Fig. \ref{fig3}, we can finally obtain the absolute 3D point cloud data. Figure \ref{fig3}(c) is the top view of the reconstructed 3D model and Fig. \ref{fig3}(d) is an oblique view for better observing the result. As known that the measured plate is used for mounting and fixing gear bearings in a mechanical watch, thus an irregularly manufactured plate will invalidate the normal function of a watch. From Fig. \ref{fig3}(c), the coplanarity and height difference between planes can be expediently checked, which provides a valid and efficient way for quality control on the production line.

\begin{figure}[htbp]
            \centerline{\includegraphics[width=1.0\columnwidth]{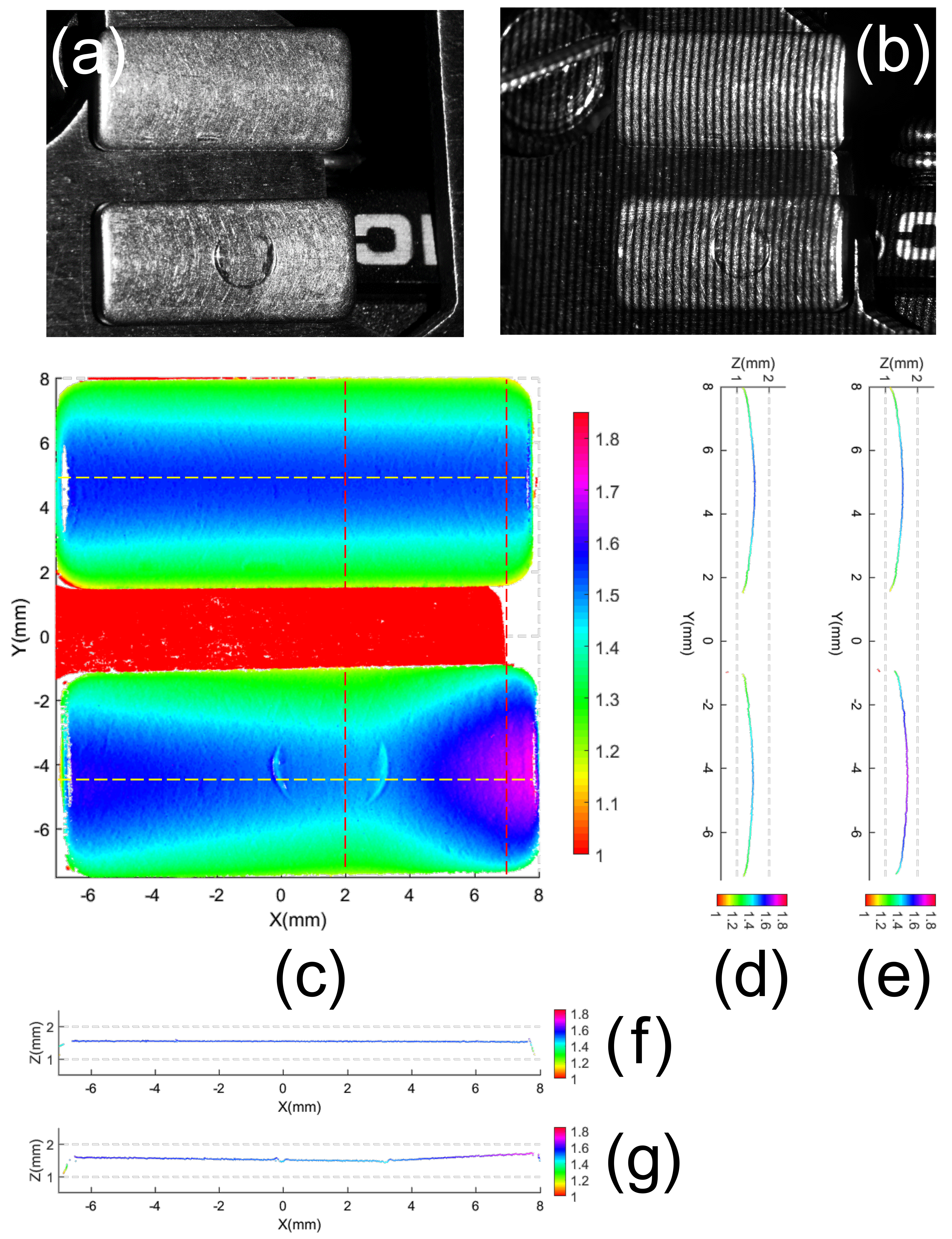}}
            \caption{
                    Experiments on the deformation measurement of two steel gaskets. (a) Sample image; (b) One of the fringe image; (c) Top view of the reconstructed 3D model; (d) and (e) Sectional views of the data as the red line labeled in (c); (f) and (g) Sectional views of the data as the yellow line labeled in (d).
            }
            \label{fig7}
         \end{figure}

        In the other experiment, the measured targets are two steel gaskets, as shown in Fig. \ref{fig7}(a). The top one has not been used to bear force and it remains in its original shape, while the other one has gotten a circle-shaped indentation after being used to decrease the pressure on the contacted surface, which can be clearly seen in Fig. \ref{fig7}(a). The measurement is to analyze the deformation of the gasket and provide quantitative information of the profile at the same time. Figure \ref{fig7}(b) is one of the fringe image, which suffers from saturation. After the same process as used in the last experiment, the absolute 3D point cloud data is obtained as shown in Fig. \ref{fig7}(c), from which we can find that the shape of the bottom gasket has been different from the top one. The indentation can be easily found from the color-coded 3D result. Due to the pressure from the weight, it is lower in the middle part while higher at both ends. Figs. \ref{fig7}(d) and (e) are two vertical sections corresponding to the red dash lines in Fig. \ref{fig7}(c). Figs. \ref{fig7}(f) and (g) are two horizontal sections corresponding to the yellow dash lines in Fig. \ref{fig7}(c). From these section views, the quantitative shape deformation of the samples can be easily acquired and used for analysis.
   \section{Conclusion} \noindent
        In this paper, we present a microscopic 3D measurement of shiny targets based on a multi-frequency phase-shifting scheme. At each pixel, only the unsaturated intensities are used to calculate the phase and the phase unwrapping process is updated by our proposed method. Considering the low depth of field in the projection light path, denser fringe patterns are easier to be defocused, and thus the integrity of the 3D result is seriously influenced. Our method tries to replace the unavailable phase at the severely saturated regions by the phase calculated from less dense fringe images. However, when dealing with those regions with constant saturation during phase-shifting, we cannot acquire helpful phases even using low-frequency fringe patterns.

        The overall process of the proposed method is detailed and the flowchart of the whole procedure is provided. The experimental results show that our method can be successfully applied in industrial applications, such as quality control and on-line inspection for micro-scale products with shiny surfaces. Actually, though the depth of field of the telecentric cameras is enough to measure samples within several millimeters, a valid solution to increase the measurable volume is using Scheimpflug principle to make the cameras have a bigger common field of view \cite{Peng15}. But the calibration of the system will become more complicated since a tilt transformation needs to be added to the imaging modeling. The future work is to further analyze how the density and phase-shifting step affect the phase accuracy when different degrees of saturation happens.

    \section*{Acknowledgments}
        \noindent
        National Key R$\&$D Program of China (2017YFF0106403), National Natural Science Fund of China (61722506, 61705105, 111574152), Final Assembly "13th Five-Year Plan" Advanced Research Project of China (30102070102), Equipment Advanced Research Fund of China (61404150202), The Key Research and Development Program of Jiangsu Province, China (BE2017162), Outstanding Youth Foundation of Jiangsu Province of China (BK20170034), National Defense Science and Technology Foundation of China (0106173), “Six Talent Peaks” project of Jiangsu Province, China (2015-DZXX-009), “333 Engineering” Research Project of Jiangsu Province, China (BRA2016407), Fundamental Research Funds for the Central Universities (30917011204, 30916011322), Open Research Fund of Jiangsu Key Laboratory of Spectral Imaging $\&$ Intelligent Sense (3091601410414), China Postdoctoral Science Foundation (2017M621747), Jiangsu Planned Projects for Postdoctoral Research Funds (1701038A).
    \section*{References}
    \bibliography{mybib}
\end{document}